\documentstyle[12pt]{article}

\newcommand{\reseteqnum}{\setcounter{equation}{0}}

\newcommand{\be}{\begin{equation}}
\newcommand{\ee}{\end{equation}}
\newcommand{\bea}{\begin{eqnarray}}
\newcommand{\eea}{\end{eqnarray}}
\newcommand{\nn}{\nonumber\\}
\newcommand{\eq}[1]{(\ref{#1})}

\catcode`\@=11
\newbox\tempboxa
\newdimen\captionboxsubcount
\def\capsize#1{\captionboxsubcount=#1pt}
\newdimen\captionboxsub
\captionboxsub=\hsize \advance\captionboxsub by -\captionboxsubcount
\advance\captionboxsub by -\captionboxsubcount
\long\def\@makecaption#1#2{
 \setbox\@tempboxa\hbox{#1: #2}
 \ifdim \wd\@tempboxa >\captionboxsub
\rightskip=\captionboxsubcount \leftskip=\captionboxsubcount #1: #2
\else \hbox to\hsize{\hfil\box\@tempboxa\hfil}
 \fi}
\catcode`\@=12
\capsize{30}

\begin{document}

\begin{titlepage}
\begin{flushright}
OU-HET 300 \\
hep-th/9807104 \\
\end{flushright}

\bigskip
\bigskip
\begin{center} \LARGE\bf
  BLACK HOLE ENTROPY \\
  FROM \\
  BPS CLOSED STRING
\end{center}
\bigskip

\begin{center} \Large
        SHUHEI MANO and YUHSUKE YOSHIDA
\end{center}
\bigskip

\begin{center} \large \it
         Department of Physics \\
         Graduate School of Science, Osaka University \\
         Machikaneyama 1-16, Toyonaka \\
         Osaka 560-0045, JAPAN \\
\end{center}

\begin{center} \Large \bf
Abstract
\end{center}
\begin{quote}
By using BPS closed string, the entropy is calculated of the extremal
five dimensional black hole consisting of Dirichlet onebranes, 
Dirichlet fivebranes and Kaluza-Klein momentum
in the flat background approximation.
In our formulations we consider two kinds of BPS closed strings with or
without a winding number.
In the former case heavy excitation modes of closed strings are used
to derive the entropy.
In the latter case we have no oscillator modes and consider collective
motion of such massless closed strings.
The entropy is given by the number of the ways how we divide the
Kaluza-Klein momentum among the massless closed strings.
In both cases the black hole entropy is the same as the
Bekenstein-Hawking entropy.
We argue that the collective modes of closed strings without
winding is equivalent to a single closed string with winding.
We propose that the two closed string pictures are connected
with the open string pictures by the modular transformation.
\end{quote}

\end{titlepage}

\section{Introduction}\label{Intro}

The recent breakthrough of string theory\cite{Polchinski} makes it
possible to calculate the Bekenstein-Hawking entropy of black holes
microscopically.\cite{Sen,SV}
The black holes are classical solution of supergravity with branes and 
the entropies are originated from brane fluctuations expressed in
terms of open strings with Dirichlet boundary condition.

Up to now we have two approaches to calculate microscopic entropy of
black holes.
One is that we consider a $\sigma$-model of the low energy effective
theory of the open string system for counting the black hole
microstates.\cite{SV,Tseytlin2,CT,BMPV,BLMPSV,HS,HMS}
This one brings us the first success of the microscopic
calculations.\cite{SV}
The other is that we consider the collective motion of massless
open strings and count the ways to divide a Kaluza-Klein
momentum among individual massless modes, giving the microscopic
entropy.\cite{CM,DM,MS}

In this paper we establish another approach by using closed strings
which are exchanged by Dirichlet branes.
We show that the black hole entropy from the closed string picture
precisely agrees with the Bekenstein-Hawking entropy.
We consider a five-dimensional extremal black hole in the type IIB
supergravity which is made of intersecting Dirichlet onebranes and
fivebranes and of Kaluza-Klein momentum along the direction of the
intersections.
The original ten-dimensional spacetime is compactified on a
fivetorus on which the branes are wrapping.
These branes exchange closed strings with a winding number.
In the low energy limit $\alpha'\to0$ the world sheets are squeezed so 
that any massive oscillator modes contribute to the black hole
entropy.
The modular transformation enable us to change this closed string
picture to the above mentioned $\sigma$-model picture of open string.
The system of closed strings will be the $\sigma$-model of 
the low energy open string effective theory.

Further, the closed string with a winding number can be regarded as a
solitonic state which consists of infinite number of closed strings
with no winding number.
The system of these closed strings with no winding number is
directly described by the type IIB supergravity.
We show that the collective motion of the closed strings have the
same physical degrees of freedom as a single closed string with
winding number.
Then, we have again the correct Bekenstein-Hawking entropy.
These closed string pictures and their interrelation with the open
string pictures will provide us with a deep understanding of the
relation  between macroscopic black hole and its microstates.

This paper is organized as follows:
First, we briefly review the black hole which is considered in this
paper in section~\ref{MBHE}.
We give the metric of the black hole and the macroscopic black hole
entropy.
The supersymmetry of the system is also discussed in the presence of
branes.
In section~\ref{BHE} we consider the BPS closed string system with
winding number, and show that we obtain the black hole entropy which
is precisely the same as the macroscopic entropy.
In section~\ref{ECS} we discuss the relation between the BPS closed
string system with a winding number and the $\sigma$-model of the low
energy effective open string theory.
We discuss the relation by using the modular transformation, and we
find that an exotic Neumann boundary condition is needed.
We derive the central charge and the target space of the closed
string system.
We conclude that the $\sigma$-model is identical with the BPS
closed string system in the low energy limit.
In section~\ref{MCS} we also derive the black hole entropy by using
infinite number of closed string with no winding.
We can regard the single closed string with a unit winding number as a 
solitonic state that consists of infinite number of closed strings
with no winding.
We discuss the relation between the closed string system with no
winding and the massless open string system mentioned above.
These system are found to be the same.
We find that the formulation of the open string system should be
modified in order to obtain correct results.
The last section is devoted to the discussions.

\section{Macroscopic Black Hole Entropy}\label{MBHE}

We work in ten-dimensional Minkowski space with the time coordinate
$x^0$ and the space coordinates $x^1, \cdots, x^9$.
The coordinates $x^5, \cdots, x^9$ are compactified on fivetorus $T^5$ 
with their radii $R_5, \cdots, R_9$.
Let us first consider $Q_1$ Dirichlet onebranes, $Q_5$ Dirichlet
fivebranes and $Q_K$ Kaluza-Klein momentum.
The Dirichlet onebranes are located at $x^1=x^2=\cdots =x^8=0$.
The Dirichlet fivebranes are located at $x^1=x^2=\cdots =x^4=0$.
We put the Kaluza-Klein momentum along the direction $x^9$.

The five dimensional extremal black hole is a solution of the type IIB 
supergravity.
The dilaton field and the ten-dimensional string metric
are\cite{Tseytlin}
\bea\label{5dbh}
e^{2\phi} &=& H_1H_5^{-1}, \nn
ds^2 &=& H_1^{1/2}H_5^{1/2}\left[
H_1^{-1}H_5^{-1}(-dx_0^2+dx_9^2 + K(dx_0-dx_9)^2) + \right.\nn
&&\left. dx_1^2 + dx_2^2 + dx_3^2 + dx_4^2 +
H_5^{-1}(dx_5^2+dx_6^2+dx_7^2+dx_8^2)\right],\nn
&&B^{RR}_{09}=\frac{1}{2}(H_1^{-1}-1),\nn
&&dB^{RR}_{ijk}=\frac{1}{2}\epsilon_{ijkl}{\partial}_l H_5,
\qquad i,j,k,l=1,\cdots,4.
\eea
The quantities $H_1$, $H_5$ and $K$ are harmonic functions in the
four-dimensional space $(x^1,x^2,x^3,x^4)$ such that
\be
H_1 = 1 + G_5\frac{4M_{D1}Q_1}{\pi r^2},\quad
H_5 = 1 + G_5\frac{4M_{D5}Q_5}{\pi r^2},\quad
K = G_5\frac{4M_{KK}Q_K}{\pi r^2},
\ee
where  $G_5 = \pi g^2\alpha'^4/(4R_5R_6R_7R_8R_9)$ is the
five-dimensional Newton constant and
$r^2 = x_1^2+x_2^2+x_3^2+x_4^2$.
$M_{D1}$, $M_{D5}$ and $M_{KK}$ are the masses of branes in the
modified Einstein frame, which are
\be
M_{D1} = \frac{R_9}{g\alpha'}, \quad
M_{D5} = \frac{R_5R_6R_7R_8R_9}{g\alpha'^3}, \quad
M_{KK} = 1/R_9.
\ee
Because of the BPS property the mass of the extreme black hole is
\be
M_{BH} = Q_1M_{D1} + Q_5M_{D5} + Q_KM_{KK}.
\ee
In order to calculate the horizon area of the black hole the metric is
sent to the Einstein frame, and the Bekenstein-Hawking entropy is
\be\label{bhe}
S_{BH} = 2\pi\sqrt{Q_KQ_1Q_5}.
\ee

Let $Q_L$ and $Q_R$ be the type IIB supercharges of positive chirality 
generated by left- and right-moving closed strings, respectively.
At the spatial infinity the system is invariant under
linear combinations $\epsilon_LQ_L+\epsilon_RQ_R$ of supersymmetries
with $\epsilon_L$ and $\epsilon_R$ are covariantly constant spinors
such that\cite{PCJ}
\be\label{susy}
\epsilon_L = \Gamma^0\Gamma^9\epsilon_R =
\Gamma^0\Gamma^5\Gamma^6\Gamma^7\Gamma^8\Gamma^9\epsilon_R =
\Gamma^0\Gamma^9\epsilon_L.
\ee
This is derived from the Killing spinor condition
\be
0 = \delta\psi_\mu = \frac{1}{\kappa}(
\partial_\mu
+ \frac{1}{4}\omega_\mu^{rs}\gamma_{rs})\varepsilon
+ \frac{ie^{-\phi}}{96}(
  H^{RR}_{\nu\rho\lambda}\gamma_\mu{}^{\nu\rho\lambda}
  - 9H^{RR}_{\mu\rho\lambda}\gamma^{\rho\lambda})\varepsilon^*
\ee
in the Schwarz's notation\cite{Schwarz} and $H^{RR}=dB^{RR}$.
The solution has the tensor product structure
\be
\epsilon_L = \epsilon_R = \xi^+_{SO(1,1)} \xi^+_{SO(4)} \xi^+_{SO(4)}, 
\ee
where $+$ means the positive chirality.
$\xi^+_{SO(1,1)}$ has one component and the two $\xi^+_{SO(4)}$ have
two components so that 4 out of the original 32 supersymmetries are
survived in the system.
The other supersymmetries become nilpotent.

The condition (\ref{susy}) tells us about the number of manifest
supersymmetry in the low energy.
What is necessary condition for considering the black hole entropy
is that in the near horizon geometry
$BTZ\times S^3\times T^4$,\cite{Hyun}
where $BTZ$ is the BTZ black hole geometry\cite{BTZ}.
So we work in the near horizon geometry.
However, we make use of the flat space approximation.
For the purpose of counting the black hole microstates we expect that
the flat space approximation will not change the number of states,
although of course other important detailed structure will change.

\section{Black Hole Entropy from BPS Closed Strings}
\label{BHE}\reseteqnum

In this section we calculate the black hole entropy in a microscopic
point of view.
We consider the above extremal black hole \eq{5dbh}.
Recent developments for calculating black hole entropy are based on the
fact that fluctuations of branes are expressed by open strings with
Dirichlet boundary condition.
In ref.~\cite{CM}
what is responsible for the entropy of the black hole is the
microscopic degrees of freedom which stem from an ensemble of
massless open strings with one end attached to a onebrane and the
other to a fivebrane.
There are two approaches to calculate the microscopic entropy.
One is that we consider a $\sigma$-model of the low energy effective
open string theory.
The other is that the entropy is given by the partition number which
is derived from counting the ways how the Kaluza-Klein momentum are
distributed among the massless open string states.

In this paper we consider a different microscopic picture from the
above two to calculate the black hole entropy.
We are interested in counting the number of BPS closed string states.
We take the weak string coupling limit $g\to0$.
We use the closed strings exchanged between the Dirichlet onebranes
and fivebranes.

For simplicity we start the discussion with the BPS closed string
exchanged by single onebrane and single fivebrane.
Suppose the closed string is wrapping once around the intersection
between the onebrane and fivebrane.
The closed string have a unit winding number in the $x^9$ direction.
\footnote{ The discussion is given in the subsection~\ref{UNITS}
why there is only the closed sting with the unit winding number.}  
We might think that the BPS closed string with winding
could not be exchanged between the onebranes and the fivebranes.
Fortunately, such closed string {\it is} exchanged,
which is discussed in subsection~\ref{BPS}.
The winding string can really propagate\cite{GKS} with a winding
number $p=\oint\gamma^{-1}\partial\gamma$ in their notation.
If the spacetime is really flat, such winding string would not be
exchanged.
We put the Kaluza-Klein momentum $n/R_9$ in the $x^9$ direction
besides winding mode.
These two data determine the zeromode of the closed string.
The zeromode part of the $X^9$ coordinate is
\be\label{closed0}
X^9 = x^9 +
\alpha'\left(\frac{n}{R_9}+\frac{R_9w}{\alpha'}\right)(\tau+\sigma) +
\alpha'\left(\frac{n}{R_9}-\frac{R_9w}{\alpha'}\right)(\tau-\sigma) +
\cdots
\ee
with $w=1$ in the flat space approximation.
The Virasoro generators, 
\bea
L_0 &=& \frac{\alpha'}{4}\left( p_\mu^2 +
    \left(\frac{n}{R_9}+\frac{R_9}{\alpha'}\right)^2\right) + N,\nn
\tilde L_0 &=&  \frac{\alpha'}{4}\left( p_\mu^2 +
    \left(\frac{n}{R_9}-\frac{R_9}{\alpha'}\right)^2\right) + \tilde N ,
\eea
vanish
where $N$ and $\tilde N$ are right- and left-moving number operator,
respectively and $\mu=0, 1, 2, 3, 4$.
We should note that the authors\cite{GKS} write down the spacetime
Virasoro generators in the NSR formalism.
For the later convenience we deal with the Green-Schwarz formalism for 
the type IIB superstring and we adopt the $(x^0,x^9)$ light-cone
gauge.
\footnote{We use the notations in the text book\cite{GSW}.}
So, the fermionic coordinates belong to the spinor representation of
$Spin(9,1)$ Lorentz group.
Let us consider the BPS string states with the five-dimensional mass
$M = n/R_9+R_9/\alpha'$.
According to the Virasoro condition the right-moving oscillator modes
freeze out, and the left-moving degrees of freedom survive;
$N=0$ and $\tilde N = N + n$.
This means that the right-mover supercharges are nilpotent;
$Q_R\approx0$.
When we consider the BPS mass in the flat space approximation,
we have the contribution $R_9/\alpha'$ from the winding mode. 
However, what we would like to consider is the near horizon geometry
$BTZ\times S^3\times T^4$, and then the extremal condition should be
$M=n/R_9$ (see for example \cite{MS2}).
In the flat space approximation this condition will be achieved
by the small radius limit $R_9\to0$.

Further, the BPS condition $\Gamma^0\Gamma^9\epsilon_L=\epsilon_L$
restricts the left-mover supercharges.
This reduces the $16=8_S+8_C$ of the original fermionic coordinates to
$8_S$.
This restriction, however, makes no effect to the physical fermionic
degrees of freedom since the restricted $8_C$ is already excluded by
the light-cone gauge $\Gamma^+\theta_L=0$.

We should notice that if we did not take any further BPS condition
into account, we had the $8_V$ bosonic and $8_S$ fermionic string
coordinates in the left-moving physical degrees of freedom.
As will be seen, the $8_V$ and $8_S$ are in fact twice as large as
the correct degrees of freedom of the BPS string with 32/8=4
supersymmetries.
We shall return to this point later, but we tentatively take $8_V$ and 
$8_S$ for a while.

Suppose that we put a onebrane and a fivebrane with a distance
$L\ne0$.
In general, the propagation of a closed string can be decomposed as a
superposition of the propagations of particles with any spin.
If the closed string propagates in the distance $L$,
not all such particles propagates in the distance $L$.
Only particles contribute this process with mass at most $M<1/L$.
This means that we have a cutoff $N_0 \sim 1/L$ in the level of the
closed string oscillator modes.
This point becomes important to explain the black hole entropy in
terms of closed string picture.
We will conclude that what is responsible for the black hole entropy
is such a closed string that propagates not through a non-zero
distance but through the zero distance.
Namely, it is the contact interaction that the closed string induces
between the onebrane and the fivebrane at their intersection.
Let us consider this point.
Let us count the degeneracy of states $d_n$ for the closed string
propagating in the distance $L$.
The massive modes with mass heavier than $1/L$ decouple in the
process.
In this situation the degeneracy of states $d_n$ is calculated from
\be
\sum_{n=0}^{\infty}d_nq^n =
\left.{\rm tr}\; q^{\tilde N}\right|_{M<1/L} =
16 \prod_{n=1}^{N_0}\left(\frac{1+q^n}{1-q^n}\right)^8.
\ee
This shows that the degeneracy of states $d_n$ is suppressed for a
large enough value of the level $n$.
So we will find it hard to produce the Bekenstein-Hawking entropy
\eq{bhe} even if we consider the ensemble of such strings.
The true situation is not this case:
Let us send $L\to0$ or equivalently $N_0\to\infty$.
Since the Dirichlet onebrane and fivebrane have the intersection,
they have the contact interaction at the intersection exchanging
the closed string.
Now, in turn, the value of $d_n$ is monotonically increasing as the
value of $n$ increases,
since any heavier modes contribute to $d_n$.
As a result, contact interaction is needed to produce arbitrary large
number of degeneracy of states.
Then, we concentrate on the closed strings which generate contact
interaction.

Next, let us consider the remaining BPS condition.
Let us count the number of states of a single closed string.
In the first a few level we have
${\rm tr}q^{\tilde N} = 16 + 256 q + 2304 q^2 + \cdots$.
These number of states are nothing but those required by the $1/4$ of
the 32 original supersymmetries.
However, the correct number of supersymmetries is just the $1/8$ of
the 32 in the present system.
The resolution is very simple to reduce more half of 8
supersymmetries.
Of course, anomaly free set of physical superstring is $8_V+8_S$.
But the supersymmetry restriction (\ref{susy}) tells us that
the relevant degrees of freedom of the low energy system is
more half BPS states of $8_V+8_S$.
The bosonic and fermionic oscillator modes $\alpha_n^i$ and $S_n^a$
should belong to a representation of the present 4 supersymmetries.
This means that they have four degrees of freedom for each level:
\be
\alpha_n^i,\quad S_n^a,\qquad i,a = 1, \cdots, 4.
\ee
Especially, if we consider the system at infinity
(or flat space approximation) the fermionic coordinates
become the spinor with the same index structure of eq.~\eq{susy}.
Now, the correct physical degrees of freedom read
\be\label{combi4}
{\rm tr}\; q^{\tilde N} =
8 \prod_{n=1}^{\infty}\left(\frac{1+q^n}{1-q^n}\right)^4 =
8 + 64q + 320q^2 + \cdots .
\ee

Next, we would like to consider the statistical ensemble,
especially the grand canonical ensemble, of such closed strings.
When we work with open strings, we consider the process that
open strings do not create and annihilate and then it is enough
to consider them as a canonical or micro-canonical ensemble.
In the case of the closed strings they are creating and annihilating.
Then, we consider them as a grand canonical ensemble.
One might think that it was strange to count the number of states of
such an intermediate closed string.
Rather, this way of counting can be seen in anywhere.
For example, similar situation is found in the system of photons
in a black box.
The number of them does not conserve and they create and annihilate
on a wall of the box.
They are also in intermediate states in a sense.

Suppose there are $Q_1$ Dirichlet onebranes and $Q_5$ Dirichlet 
fivebranes.
Since the two kinds of branes have $Q_1Q_5$ intersections, we have
$Q_1Q_5$ closed strings.
\footnote{Instead of considering the present case in which all
onebranes and fivebranes have unit winding number,
we may consider another cases in which, for example, 
single onebrane has winding number $Q_1$ and single fivebrane has 
winding number $Q_5$. This is discussed in the subsection~\ref{UNITS}}
Each closed string has a Kaluza-Klein momentum $n_i$ with a given
total Kaluza-Klein momentum; $Q_K = \sum_{i=1}^{Q_1Q_5}n_i$.
We have a degeneracy of dividing the total number $Q_K$ into the
positive integers $n_i$ ($i=1,2,\cdots,Q_1Q_5$) of the individual
closed strings.
Taking this degeneracy into account, we obtain
\be\label{combi}
\sum_{n=0}^{\infty}d_nq^n =
\left({\rm tr}\; q^{\tilde N}\right)^{Q_1Q_5} \approx
\exp\left(-\frac{\pi^2}{6}\frac{c}{\ln q}\right),\qquad (q\nearrow1) ,
\ee
where we have a central charge $c = 4(1+1/2)Q_1Q_5$.
In the leading order a simple saddle point evaluation allows us to
obtain
\be\label{dos}
d_n = \exp \left(2\pi\sqrt{\frac{nc}{6}}\right).
\ee
The degeneracy of states \eq{dos} with $n=Q_K$ and the center
$c=4(1+1/2)Q_1Q_5$ precisely produce the Bekenstein-Hawking
entropy \eq{bhe}.

Finally, we consider the mass of the present black hole.
Adding up the masses of onebranes, fivebranes and BPS strings,
we have the total mass
\be
M_{BH} = Q_1M_{D1} + Q_5M_{D5} + Q_KM_{KK}.
\ee
We have to care about a contribution from the winding mode of
the BPS strings.
In the flat space the black hole mass has the contribution
$R_9Q_1Q_5/\alpha'$ from the winding mode.
However, the near horizon geometry do not produce such a contribution
as is already explained.
In any case if we consider the weak string coupling limit $g\to0$ and
the small radius limit $R_9\to0$,
the mass of winding mode becomes negligible compared with the masses
of onebrane, fivebrane and Kaluza-Klein momentum.

\subsection{Units of the winding number}
\label{UNITS}\reseteqnum

In this subsection we discuss that there are several choices
of the units of winding numbers and why there are only
the closed strings with the unit winding number.

In this paper we consider $Q_1$ onebranes and $Q_5$ fivebranes with
winding number one around the $x^9$ direction with radius $R_9$.
Instead of this case there are many other possibilities giving the
same result.
Let us suppose that there are $q_1$ onebranes with a winding number
$w_1$ and $q_5$ fivebranes with a winding number $w_5$.
Then, the total Ramond-Ramond charges are $Q_i= w_iq_i$ ($i=1,5$),
respectively, for onebranes and fivebranes.
We assume that $w_1$ and $w_5$ are relatively prime number.
In the following discussion we concentrate on a specific onebrane and
a specific fivebrane.

Let us consider the process that a closed string is emitted from the
onebrane and is absorbed into the fivebrane.
Since the closed string has to wrap around the onebrane integral times,
the closed string has to have winding number in unit of $w_1$; $w =
w_1n$ where $n\in\mbox{\bf Z}$.
On the other hand, the same argument should hold for the fivebrane, 
and thus the closed string also has to have winding number in unit of
$w_5$.
In order to satisfy the above two conditions, we find the closed
string winding number $w = w_1w_5m$ where $m\in\mbox{\bf Z}$.

In this situation we put a Kaluza-Klein momentum $p_9=n/R_9$ on the
closed string.
The BPS mass of the closed string is $M = n/R_9 + mw_1w_5R_9/\alpha'$.
Since the mass is linear in $n$ and $m$, this closed string can split
into $m$ pieces with unit winding number $w_1w_5$ and mass
$M_i = n_i/R_9 + w_1w_5 R_9/\alpha'$ where $n_1+n_2+\cdots+n_m = n$.
Then, it will be enough to consider the closed string with unit
winding number $w_1w_5$.

In this case the Virasoro condition of the total system is $N=0$ and
$\tilde{N}=w_1w_5Q_K$.
Since we are considering $q_1q_5$ such closed string, the total
central charge is $c = 6q_1q_5$ and the entropy is
$S = 2\pi\sqrt{\tilde{N}c/6} = 2\pi\sqrt{Q_1Q_5Q_K}$.
This final formula does not depend on the ways how we divide $Q_i$
into two integers $q_i$ and $w_i$.

\subsection{BPS property and exchanging winding strings}
\label{BPS}\reseteqnum

In this subsection we show that
in the flat background approximation the BPS closed string
with a winding number can be exchanged by the onebranes and
the fivebranes in the weak string coupling and small radius limits;
$g \to 0$ and $R_9 \to 0$. In fact it is seen that the number of global 
supersymmetries is the same as that of the system without 
a winding number.
We notice that winding string indeed can be exchanged in the
near horizon geometry\cite{GKS}.

Since the ten-dimensional spacetime is compactified by a fivetorus to 
a five-dimensional spacetime, onebranes and fivebranes are looked as a 
point particle sitting at the origin seen from the five-dimensional
spacetime.
The winding number of the fundamental string is the conserved charge
of the gauge symmetry stemming from the NS-NS two-form field.
Suppose that the onebranes and fivebranes exchange a closed string
with a winding number.
No matter how onebranes and fivebranes change their NS-NS charge from
zero, the total NS-NS charge at the origin is conserved and is zero.
Then, we have no NS-NS gauge field emerging from the origin.
Then, the supergravity solution of this system is identical with the
ordinary one for Dirichlet onebrane, Dirichlet fivebrane and
Kaluza-Klein momentum.
This is an important fact to explain why the black hole has
a huge number of degeneracy of its microstates.

Next, let us see that in the flat background approximation
it is possible to exchange winding string and
the four global supersymmetries survive.
Since we are taking the weak string coupling limit $g\to0$, the charge
lattice of NS-NS and R-R two form fields collapses.
We are also taking the small radius limit $R_9\to0$.
The mass of the winding string is negligible compared with the mass of
the branes.
Thus, branes can emit and absorb winding string without energy.
Then, after emitting and absorbing winding string, the system is still 
BPS state.
In order to make sure this point we show two discussions to find that
we have four global supersymmetries.
First, we start the discussion with open string with Dirichlet
boundary condition.
According to the standard discussion\cite{Polchinski} the open string
has four global supersymmetries.
Now, suppose that this open string propagates around the compact $x^9$
space making a world sheet which can be interpreted as that of closed
string with a winding number.
This closed string is nothing but what we are dealing with.
Then, we have four global supersymmetries.
This result also can be obtained using supergravity.
As is recognized above, the supergravity solution of the system with
intermediating winding strings is identical to that without winding 
strings.
The authors\cite{GKS} also considered the winding string with
winding number, $p=\oint\gamma^{-1}\partial\gamma$ in their notation,
in the same background $AdS_3\times S^3\times T^4$.
These two solution provide us with the same condition for the global
supersymmetry; $\delta\psi_\mu=0$, and thus the two system have the
same global supersymmetry.

\section{Ensemble of BPS Closed Strings as a $\sigma$-model}
\label{ECS}\reseteqnum

In this section let us discuss about the relation between open
and closed string pictures.
Before doing this we recall that the brane fluctuations are expressed
in terms of open strings with Dirichlet boundary condition.\cite{CM}
We will find it convenient to briefly recall the quantization of
branes by open strings.
There are three types of open strings which are connecting two
onebranes, two fivebranes and a onebrane at one end and a fivebrane at 
the other end.
We call these (1,1), (5,5) and (1,5) strings, respectively.
In the following we consider onebrane worldvolume gauge theory,
although fivebrane worldvolume theory will give the same results.
There are $Q_1Q_5$ different (1,5) strings since we have $Q_1$
onebranes and $Q_5$ fivebranes.
The (1,5) strings form a matter with the bi-fundamental representation
in the $U(Q_1)$ and $U(Q_5)$ gauge theories.
The degrees of freedom of the (1,1) and (5,5) strings are dropped by
$D$-flatness and gauge fixing conditions.\cite{Maldacena}
The relevant degrees of freedom come from (1,5) strings.
In summary, the physical fluctuations of onebranes and fivebranes are
described by $Q_1Q_5$ (1,5) strings in the onebrane worldvolume
gauge theory.

Now let us start the discussion with the simplest case $Q_1=Q_5=1$.
The relevant (1,5) open string has the boundaries on the onebrane at
one end and on the fivebrane at the other end.
As increasing the world sheet time $\tau$ from $0$ to $\pi$,
\footnote{The cylinder amplitude has modulus. In the following
discussion, we use the particular value of the modulus.}
the boundary point of the (1,5) open string on the onebrane
sweeps along the onebrane and the boundary point on the other end
draws a copy of a circle on the fivebrane.
Interchanging the parameters $\sigma\to\tau$ and $\tau\to-\sigma$, the
open string world sheet can be interpreted as the world sheet of a BPS
closed string.\cite{Polchinski}
Notice that the time $\tau$ of the closed string is corresponding
to $\sigma$ of a open string and vise versa.
The argument by Polchinski \cite{Polchinski} 
should be naturally extended to the 
case with both the Kaluza-Klein and winding modes.
Namely, the closed string world sheet is rewritten in terms of the
above (1,5) open string by the modular transformation.
We know that in terms of the closed string picture the onebrane and
the fivebrane become the boundary states of a closed string at
$\tau=0,\pi$ with winding number $w=1$ in the $x^9$ direction.
Then, corresponding to this fact, the open string should have a
winding number around its $\tau$ direction.
The resultant world sheet is topologically a copy of an annulus.
For definiteness we consider that the inner boundary circle of the
annulus is on the onebrane and the outer one is on the fivebrane.

We think that this closed/open string correspondence needs more
explanation:
The closed/open correspondence by the modular transformation should be 
persisted in not only in the oscillator modes but also in the
zeromodes.
For the purpose of explaining, we begin with a BPS closed string
in the flat background approximation.
If the BPS closed string has both the Kaluza-Klein momentum $n/R_9$
and the winding number $R_9w/\alpha'$, the corresponding open string
also does in some sense.
Here $n,w = 0, \pm1, \pm2, \cdots$.
We impose the boundary condition of BPS closed string and
interchanging with $\tau\to-\sigma$ and $\sigma\to\tau$.
The resultant open string coordinate is
\be\label{open}
X^9 = x^9 +
\alpha'\left(\frac{n}{R_9}+\frac{R_9w}{\alpha'}\right)(-\sigma+\tau) +
\alpha'\left(\frac{n}{R_9}-\frac{R_9w}{\alpha'}\right)(-\sigma-\tau) +
\cdots.
\ee
This open string has, in fact, an exotic Neumann boundary condition.
How should we modify the conventional argument of the open string
boundary condition?
Let us consider this point.
Let $S_0$ be the Polyakov action for the open string.
The variation of the action is
\be\label{delS0}
\delta S_0 =
- \frac{1}{2\pi\alpha'}\int d^2\sigma \delta
X_\mu\partial^\alpha\partial_\alpha X^\mu
+
\frac{1}{2\pi\alpha'}\int d\tau
\left[\delta X^\mu\partial_\sigma X_\mu
      \right]_{\sigma=0}^{\sigma=\pi}.
\ee
The first term gives the equation of motion.
Substituting the solution \eq{open} into eq.~\eq{delS0}, however, the
second term does not annihilate.
To save the situation we have to consider that the open string with
winding number should have a interaction with the boundary.
All we have to do is to find such an interaction that cancels
the contribution from the boundary in eq.~\eq{delS0}.
The desirable interaction is
\be
S_{int} = \frac{n}{\pi R_9}\int d\tau
\left. X^9 \right|_{\sigma=0}^{\sigma=\pi}.
\ee
Now, the total action $S=S_0+S_{int}$ has the correct variation.
We conclude that the closed string Kaluza-Klein momentum is
interpreted as a winding mode of the open string with the exotic
Neumann boundary condition.
 
We explain why there is a one-to-one relation between the
degrees of freedom of the relative motion of the D1, D5 branes and
that of a single closed string.
For counting the relative degrees of freedom of D1-D5 brane system,
we consider not many closed strings but a single closed string.
We would like to discuss this point.
It is well known that the relative motion between the onebrane and 
the fivebrane is described by a single open string stretching between 
them. 
In our picture, the worldsheet of this open string can be interpreted 
as that of a closed string.
Then, the relative motion corresponds to the single closed string.
We also consider the above discussion in terms of the closed string
picture in the following way:
Simultaneously D-brane does not emits or absorbs many closed sting
because of the fact that D-branes have a unit Ramond-Ramond charge.
Suppose the branes exchange $m \ne 1$ closed strings.
Notice that, as shown in the subsection~\ref{UNITS},
even though we consider the single closed string with
winding number $m$ decays into $m$ closed strings with unit winding
number.
So, it is enough to consider closed strings
with unit winding number.
Each closed string with unit winding number contribute to the evaluation
of the Ramond-Ramond charge, as was demonstrated by
Polchinski\cite{Polchinski}, by a unit charge.
Then, the net Ramond-Ramond charge is $m$, which contradicts the fact
that brane has a unit Ramond-Ramond charge.
Therefore, we consider a single closed string with unit winding number.

Here we come back to the discussion.
We generalize the above closed/open correspondence to the cases
when we have many onebranes and fivebranes $Q_1,Q_5>1$.
In this case the open strings have Chan-Paton factor.
This factor indicates which onebrane and fivebrane the open strings
connect with.
In terms of closed string language, we can say that the Chan-Paton
factor is interpreted as some kind of index
which onebrane and fivebrane the closed string begins to propagate from
and terminates at.
Then, we have $Q_1Q_5$ closed string world sheets.
In conclusion, the situation is summarized as follows:
The Chan-Paton factor is not a dynamical object but we have $Q_1Q_5$
world sheets.
So, we have $Q_1Q_5$ closed string world sheets which are
topologically a copy of an annulus.

Now, let us consider the low energy limit $\alpha'\to0$.
The limit $\alpha'\to0$ forces the area of the world sheets to be
vanishing by a familiar energetic argument.
Namely, the world sheet is squeezed.
The world sheets become a copy of a circle, and does not shrink to a
point since it has a unit winding number.
Since open string boundaries are restricted on the onebranes and the
fivebranes, the world sheets are restricted on their intersections.
In terms of open string language, in the $\alpha'\to0$ limit the
lowest lying modes dominate in the spectrum, which are
hypermultiplet matter in a gauge field theory.\cite{Witten,Witten2}
While in terms of closed string language massive modes does not
decouple because of the reason explained in section \ref{BHE}.
Squeezed worldsheet of a closed string becomes a worldline of
a hypermultiplet matter of a (1,5) string in the onebrane worldvolume
gauge theory.
Considering the physical degrees of freedom in the BPS closed string
system as in section \ref{BHE}, we have a center
\be\label{c}
c = 4\left(1+\frac{1}{2}\right)Q_1Q_5.
\ee

Finally, we show that the target space of the closed string system is
$(T^4)^{Q_1Q_5}/S_{Q_1Q_5}$ where $S_{Q_1Q_5}$ is the permutation
group on $Q_1Q_5$ objects.
Since the world sheet of BPS closed string is interpreted by that of
open string, through the modular transformation, it is sufficient to
start with considering using open string.
Suppose the onebranes are fluctuating in the $(x^5, x^6, x^7, x^8)$
directions.
Let $\phi_i$ ($i=5,6,7,8$) be the position of a onebrane in the
$(x^5, x^6, x^7, x^8)$ directions.
The position in the non-compact directions is irrelevant here.
During the onebrane fluctuations the (1,5) open strings do not acquire
mass from the string tension, since the fivebranes have the
worldvolume along the $(x^5, x^6, x^7, x^8)$ directions.
Namely, the onebranes and the fivebranes are always intersecting
during the fluctuations.
In order to describe such fluctuations we consider the vertex
operator\cite{PCJ} for the position of onebrane which is
$V=\phi_i(X)\partial_\sigma X^i$ ($i=5,6,7,8$).
This shows that the massless fluctuations $\delta\phi_i$ of onebranes
are driven by the open string coordinates $X^i$ ($i=5,6,7,8$).
This result is easily interpreted in terms of closed string language:
The brane fluctuations preserving contact closed string interaction is 
driven by the closed string coordinates $X^i$ ($i=5,6,7,8$).
Then, the target space of a single closed string is described by
$X^i$ ($i=5,6,7,8$) which is a position in the fourtorus $T^4$.
Considering all the closed strings, the total target space would be
$(T^4)^{Q_1Q_5}$.
Since we have multiplicity in the onebrane and fivebrane intersections
as well as self-intersections the target space should be divided by
the symmetry factor $S_{Q_1Q_5}$.
Then, we have the desired target space\cite{Vafa}
\be\label{target}
(T^4)^{Q_1Q_5}/S_{Q_1Q_5}.
\ee
In conclusion, the central charge \eq{c} and target space \eq{target}
precisely agree with the result of the $\sigma$-model\cite{SV,CM} from
open string.
The present closed string system is identical to the
$\sigma$-model\cite{SV,CM} of the low energy effective open string
system.

\section{Black Hole Entropy from Massless Closed Strings}
\label{MCS}\reseteqnum

In this section, we re-derive the black hole entropy microscopically
by using the infinite number of closed strings with no winding.

For simplicity we start with considering a single closed string with
no winding and a onebrane and a fivebrane.
The closed string is exchanged between the onebrane
and the fivebrane with the contact interaction.
We put Kaluza-Klein momentum $n/R_{9}$ along the $x^9$ 
direction on this closed string where $n$ is a positive integer.
Then, the closed string is a massless BPS state with $p_0=n/R_9$.
\footnote{In the section, the word ``massless'' is used in 
ten dimensions.} 
In the $\alpha'\to0$ limit, the worldsheet shrink to a worldline since
there is no winding number.
The relevant degrees of freedom are center-of-mass motions.
Then, the closed string has no oscillator modes; $N = \tilde{N} = 0$.
The situation is the same as that in ref.~\cite{DM}
if their open string is replaced to the closed string.
The number of the supercharges is the same as that in
eq.~\eq{susy}, and we have four supercharges.
The four supersymmetries implies that we have four bosonic and
fermionic ground states, denoted by $|n,i\rangle$ and
$|n,\dot{a}\rangle$ ($i,\dot{a} = 1,\cdots,4$), respectively.
For a while, we consider the number of brane intersections as
$n^\sharp=1$.

Next, we consider the ensemble of the above strings.
Let $N_{n,i}^b$ and $N_{n,\dot{a}}^f$ be the number of the bosonic
$|n,i\rangle$ and fermionic $|n,\dot{a}\rangle$ states, respectively.
The total Kaluza-Klein momentum is
\be\label{totalKK}
\sum_{n=1}^{\infty} \sum_{i=1}^{4n^\sharp}
\frac{n}{R_9}\left( N_{n,i}^b + N_{n,i}^f\right) ,
\ee
which should be identical to $Q_K/R_9$.
There are many ways to divide the total Kaluza-Klein momentum
$Q_K/R_9$ among the individual massless BPS states.
The number of the ways is
\be\label{nos1}
d_{Q_K} =
8^{n^\sharp}\left(\prod_{n=1}^\infty\prod_{i=1}^{4n^\sharp}
\sum_{N_{n,i}^b=0}^\infty \sum_{N_{n,i}^f=0}^1 \right)
\delta \left( Q_K,\sum_{n=1}^\infty \sum_{i=1}^{4n^\sharp} n
\left(N_{n,i}^b + N_{n,i}^f\right) \right) ,
\ee
where $\delta(m,n)$ is the Kronecker delta.
Here we multiply by the degeneracy $8^{n^\sharp}$ of the ground states
with no Kaluza-Klein momentum; $|0,i\rangle$ and $|0,\dot{a}\rangle$.
We notice that in the above equation we have infinite number of
summations over all integer variables $N_{n,i}^b$ and $N_{n,i}^f$
($n=1,2,\cdots,\infty$ and $i,\dot{a}=1,\cdots,4$).
Then, we obtain
\bea\label{nosq}
\sum_{Q_K=0}^{\infty}d_{Q_K}q^{Q_K}
&=&
   8^{n^\sharp}\prod_{n=1}^\infty\prod_{i=1}^{4n^\sharp} \left(
     \sum_{N_{n,i}^b=0}^\infty \sum_{N_{n,i}^f=0}^1
     q^{n \left(N_{n,i}^b + N_{n,i}^f\right)}
   \right)\nn
&=&
   8^{n^\sharp}\prod_{n=1}^{\infty}
   \left(\frac{1+q^{n}}{1-q^{n}}\right)^{4n^\sharp}.
\eea
This result shows that the infinite number of closed strings with no
winding have the same degrees of freedom as that, in eq.~\eq{combi4}, 
of the single closed string with unit winding number.

Now, we consider the general case $Q_1,Q_5>1$.
There are $Q_{1}Q_{5}$ intersections, and we have exactly the same
closed string system as the above one in each intersection.
In this case the total number of states $d_{Q_K}$ is given by
eq.~\eq{nos1} with $n^\sharp=Q_1Q_5$.
The number of states $d_{Q_K}$ in the large $Q_K$ limit is easily
calculated from eq.~\eq{nosq} with $n^\sharp=Q_1Q_5$, and we obtain the
entropy
\be\label{s}
S = \ln d_{Q_K} = 2\pi\sqrt{\frac{Q_Kc}{6}} ,
\ee
 where $c = 4(1+1/2)Q_{1}Q_{5}$.
We derive the black hole entropy microscopically by using the infinite
number of closed strings with no winding.
The entropy is precisely the same as that calculated from
eq.~\eq{combi} with $c=6Q_1Q_5$ which was derived from the 
closed string system with a winding number in section~\ref{BHE}.

In the above calculations we find three facts:
As seen from eq.~\eq{nosq} the two closed string systems with winding
and without winding have exactly the same degrees of freedom.
The two systems give the same microscopic entropy of the black hole.
Let us consider the small radius limit $R_9\to0$.
Even in the flat background approximation the two closed string system
have the same five-dimensional mass $M=Q_K/R_9$ in generic values of
$Q_1$ and $Q_5$.
Thus, we can regard the single closed string with a unit winding
number as a solitonic state that consists of infinite number of closed
strings with no winding.

Next, we interpret the present picture of closed strings with no
winding as that of open strings.
Interchanging the parameters $\sigma \rightarrow -\tau$ and 
$\tau \rightarrow \sigma$, the closed string world sheet 
is converted into the open string world sheet as explained in
section~\ref{ECS}.
The zeromode part of the resultant open string $X^9$ coordinate is
\be\label{w=0}
X^9 =x^9 + 2 \alpha'\frac{n}{R_9}\sigma + \cdots .
\ee
In this open string picture, the original Kaluza-Klein momentum $n/R_9$
looks as if it appears as a winding number along the $x^9$ direction
with the radius $R_9'=\alpha'/R_9$.
If the radius of the $x^9$ direction is $R_9'$, the open string
will satisfy the periodic boundary condition
$X^9(\sigma+\pi)=X^9(\sigma)+2\pi nR_9'$ and will looks like a closed
string.
What the original Kaluza-Klein momentum is a good quantum number
is explained by the fact that such ``closed string'' has the winding
number $n$ along the $x^9$ direction with the radius $R_9'$.
On the other hand, this open string has no winding along the $x^9$
direction, since it is related to the closed string with no winding
by the modular transformation.

Finally, we comment on the work in refs.~\cite{CM,DM} using the open
string picture mentioned in the beginning of section~\ref{BHE}.
The zeromode part of the $X^9$ coordinate%
\footnote{
Here we rewrite the $X^5$ in refs.~\cite{CM} into our notation
$X^9$.}
of open strings in refs.~\cite{CM,DM} is
\be\label{openzero}
X^{9} = x^{9} + 2 \alpha' \frac{n}{R_{9}} \tau + \cdots .
\ee
Since the $x^9$ direction is compactified with the radius $R_9$,
we need a periodic boundary condition on the string coordinate
\eq{openzero}.
However, the string coordinate jumps by $2\pi n\alpha'/R_9$ when one
circles along the $\tau$ direction in the world sheet.
This condition contradicts the periodic boundary condition.
Further, without winding along the $x^9$ direction, it is seemed that
the Kaluza-Klein momentum of the open strings cannot be a good quantum 
number.
Therefore, we should modify eq.~\eq{openzero} to eq.~\eq{w=0}.

\section{Discussions}\reseteqnum

First, let us discuss the relations between several approaches to
calculate the black hole entropy.
Let us consider the BPS closed string system with no winding number.
In the flat background, the presence of branes leads to a situation
that the worldsheet boundary may be regarded as a kind of non-vanishing
tadpoles of closed string.
If we sum over all tadpoles, we can recover the true vacuum with
vanishing tadpole.
The condition of vanishing tadpole is satisfied by the conformal
invariance of the system.
Namely, the system of the BPS closed string with no winding and with
branes is described by type IIB supergravity.
In conclusion, the dynamics of BPS closed strings with no winding is
governed by the conventional type IIB supergravity.
We notice that we do not take the limit $\alpha'\to0$.
This simple picture provides us with the equivalence between the
microscopic and macroscopic black hole entropies.
Further, as discussed in the present paper, the closed string with
winding number is a solitonic state of the system of the BPS closed
strings with no winding.
Then, the three theories should be mutually equivalent.
The BPS closed strings with a winding number has no such correspondence
as that between massless BPS strings and supergravity,
or rather it becomes a kind of $\sigma$-model from open string system
proposed by ref.~\cite{SV}.

Finally, let us discuss, in non-extremal case, the expression of the
entropy
\be\label{entropy}
S = 2\pi(\sqrt{n_L}+\sqrt{n_R})
(\sqrt{Q_1}+\sqrt{Q_{\bar 1}})(\sqrt{Q_5}+\sqrt{Q_{\bar 5}}) ,
\ee
taking the discussion in ref.~\cite{HMS} into consideration.
If we change branes to its anti-branes or left-moving Kaluza-Klein
momentum to right-moving one in the extremal system \eq{5dbh}, we have
eight kinds of BPS states.
If we consider the non-extremal system with
$Q_1,Q_{\bar 1},Q_5,Q_{\bar 5},n_L,n_R \ne 0$,
the relevant string coordinate is given by the eight individual BPS
string coordinates.
Then, the entropy formula \eq{entropy} is looked as if it is saturated
by the contributions from the eight individual minimal BPS states.
We have the correct black hole mass formula
\be
M_{BH} = (Q_1+Q_{\bar 1})M_{D1} + (Q_5+Q_{\bar 5})M_{D5} +
(n_L+n_R)M_{KK}
\ee
in contrast with the
discussion in ref.~\cite{HMS}.

To simplify the discussion let us first consider the case in which we
have eight supersymmetries.
If we include Kaluza-Klein momentum in the left and right sectors,
the number of supersymmetries increases by twice compared with the
extremal case in the present paper.
While we already have the four supersymmetries $4_L$ in the
left-moving sectors, the four supersymmetries $4_R$ are restored in
the right-moving sectors; $8 = 4_L + 4_R$.
Each BPS state of left-mover and right-mover contains four bosonic and
fermionic oscillator modes as explained in section~\ref{BHE}.
Let us call this BPS state minimal BPS state.
Fortunately, the multiplets of this eight supersymmetries are BPS
states of  the original 32 IIB supersymmetries.
Then, thanks to the BPS property, we can carry out a reliable
calculation perturbatively for deriving the entropy.
In the lowest string coupling order the left and right sectors are
calculated independently.
Namely, the entropy is saturated by the two individual minimal BPS
states, and we obtain
\be\label{nonextremal}
S = 2\pi(\sqrt{n_L}+\sqrt{n_R})\sqrt{Q_1Q_5} .
\ee
This result is converted, by $U$-dualities\cite{Hyun,CM,HMS},
to the other two cases in which we have anti-onebranes or
anti-fivebranes in addition to onebranes, fivebranes and
left-moving Kaluza-Klein momentum.
Then, we can carry out reliable calculations for deriving the three
entropies \eq{nonextremal} with both $(n_L,Q_1,Q_5)$ and
$(n_R,Q_{\bar 1},Q_{\bar 5})$ cyclically permuted.
The permutations are induced by $U$-duality.
Moreover, we can generalize the system to that with one-, five-
branes and their anti-branes and left-moving Kaluza-Klein momentum.
This system is still BPS state of the original IIB string theory.
Taking 4 individual minimal BPS states into account, we obtain the
entropy
\be
S = 2\pi\sqrt{n_L}(\sqrt{Q_1}+\sqrt{Q_{\bar1}})
(\sqrt{Q_5}+\sqrt{Q_{\bar5}}) .
\ee
This entropy can be converted to other two cases, as the above
\eq{nonextremal} case, by $U$-duality.

What we learn in the above is that we have eight mutually independent
minimal BPS states in the weak string coupling limit $g\to0$.
Since the number of BPS states hopefully does not change under the
variation of the value of the string coupling constant, we
microscopically obtain the result \eq{entropy}.


\begin{flushleft}
\Large Acknowledgments 
\end{flushleft}
The work was supported in part by the Research Fellowships of the
Japan Society for the Promotion of Science for Young Scientists.



\begin{thebibliography}{99}

\bibitem{Polchinski}
  J. Polchinski,
  {\it Phys. Rev. Lett.} {\bf 75} (1995) 4724, hep-th/9510017,
  ``DIRICHLET BRANES AND RAMOND-RAMOND CHARGES''.

\bibitem{Sen}
  A. Sen,
  {\it Mod. Phys. Lett.} {\bf A10} (1995) 2081, hep-th/9504147,
  ``EXTREMAL BLACK HOLES AND ELEMENTARY STRING STATES''.

\bibitem{SV}
  A. Strominger and C. Vafa,
  {\it Phys. Lett.} {\bf B379} (1996) 99, hep-th/9601029,
  ``Microscopic Origin of the Bekenstein-Hawking Entropy''.

\bibitem{Tseytlin2}
  A.A. Tseytlin,
  {\it Nucl. Phys.} {\bf B477} (1996) 431, hep-th/9605091,
  ``Extremal black hole entropy from conformal string sigma model''.

\bibitem{CT}
  M. Cveti\v{c} and A.A.Tseytlin,
  hep-th/9806141,
  ``Sigma Model of Near-Extreme Rotating Black Holes and Their Microstates''.

\bibitem{BMPV}
  J.C. Breckenridge, R.C. Myers, A.W. Peet and C. Vafa,
  {\it Phys. Lett.} {\bf B391} (1997) 93, hep-th/9602065,
  ``D-branes and Spinning Black Holes''.

\bibitem{BLMPSV}
  J.C. Breckenridge, D.A. Lowe, R.C. Myers, A.W.Peet, A. Strominger 
  and C. Vafa,
  {\it Phys. Lett.} {\bf B381} (1996) 423, hep-th/9603078,
  ``MACROSCOPIC AND MICROSCOPIC ENTROPY OF NEAR-EXTREMAL 
  SPINNING BLACK HOLES''.

\bibitem{HS}
  G.T. Horowitz and A. Strominger,
  {\it Phys. Rev. Lett.} {\bf 77} (1996) 2368, hep-th/9602051,
  ``Counting States of Near-Extremal Black Holes''.

\bibitem{HMS}
  G.T. Horowitz, J.M. Maldacena and A. Strominger,
  {\it Phys.Lett.} {\bf B383} (1996) 151, hep-th/9603109,
  ``Nonextremal Black Hole Microstates and U-Duality''.

\bibitem{CM}
  C.G. Callan and J.M. Maldacena,
  {\it Nucl.Phys.} {\bf B472} (1996) 591, hep-th/9602043,
  ``D-Brane Approach to Black Hole Quantum Mechanics''.

\bibitem{DM}
  S.R. Das and S.D. Mathur,
  {\it Phys.Lett.} {\bf B375} (1996) 103, hep-th/9601152,
  ``Excitations of D-strings, Entropy and Duality''.

\bibitem{MS}
  J.M. Maldacena and A. Strominger,
  {\it Phys. Rev. Lett.} {\bf 77} (1996) 428, hep-th/9603060,
  ``Statistical Entropy of Four-Dimensional Extremal Black Holes''.

\bibitem{Tseytlin}
  A.A. Tseytlin,
  {\it Nucl. Phys.} {\bf B475} (1996) 149, hep-th/9604035,
  ``Harmonic superpositions of M-branes''.

\bibitem{PCJ}
  J. Polchinski, S. Chaudhuri and C.V. Johnson,
  NSF-ITP-96-003, hep-th/9602052,
  ``Notes on D-Branes''.

\bibitem{Schwarz}
  J.H. Schwarz,
  {\it Nucl. Phys.} {\bf B226} (1983) 269,
  ``COVARIANT FIELD EQUATIONS OF CHIRAL N=2 D=10 SUPERGRAVITY''

\bibitem{BTZ}
  M. Ba$\tilde{\rm n}$ados, C. Teitelboim and J. Zanelli,
  {\it Phys. Rev. Lett.} {\bf 69} (1992) 1849,
  ``Black Hole in Three-Dimensional Spacetime''.

\bibitem{GKS}
  A. Giveon, D.Kutasov and N. Seiberg,
  {\it Adv. Theor. Math. Phys.} {\bf 2} (1998) 733,
  ``Comments on String Theory on $AdS_3$''.

\bibitem{GSW}
  M.B. Green, J.H. Schwarz and E. Witten,
  Cambridge Univ. press. 1987,
  ``Superstring theory''.

\bibitem{MS2}
  J. Maldacena and A. Strominger,
  {\it JHEP} {\bf 005} (1998) 9812, hep-th/9804085,
  ``$AdS_3$ Black Holes and a Stringy Exclusion Principle''.

\bibitem{Maldacena}
  J.M. Maldacena,
  Ph. D. Thesis, Princeton University, June 1996, hep-th/9607235,
  ``Black Holes in String Theory''.

\bibitem{Witten}
  E. Witten,
  {\it Nucl. Phys.} {\bf B460} (1996) 335, hep-th/9510135,
  ``BOUND STATES OF STRINGS AND P-BRANES''.

\bibitem{Witten2}
  E. Witten,
  {\it JHEP} {\bf 003} (1997) 9707, hep-th/9707093,
  ``ON THE CONFORMAL FIELD THEORY OF THE HIGGS BRANCH''.

\bibitem{Vafa}
  C. Vafa,
  {\it Nucl. Phys.} {\bf B463} (1996) 415, hep-th/9511088,
  ``Gas of D-Branes and Hagedorn Density of BPS States''.

\bibitem{Hyun}
  S. Hyun, 
  hep-th/9704005,
  ``U-duality between Three and Higher Dimensional Black Holes''.


\end{thebibliography}
\end{document}